\documentclass[ reprint, superscriptaddress, amsmath, amssymb, aps, prb]{revtex4-1}

\usepackage{graphicx}   
\usepackage{dcolumn}    
\usepackage{bm}         
\usepackage{CJK}

\begin{document}

\title{High-precision Monte Carlo study of  
the  three-dimensional XY model on GPU}   
\begin{CJK}{UTF8}{bsmi}
\author{Ti-Yen Lan (籃棣彥)}
\affiliation{Center of Theoretical Science and Department of Physics, National Taiwan University, Taipei 10607, Taiwan}

\author{Yun-Da Hsieh (謝昀達)}
\affiliation{Center of Theoretical Science and Department of Physics, National Taiwan University, Taipei 10607, Taiwan}

\author{Ying-Jer Kao (高英哲)}
\email{yjkao@phys.ntu.edu.tw}
\affiliation{Center of Theoretical Science and Department of Physics, National Taiwan University, Taipei 10607, Taiwan}
\affiliation{Center for  Advanced Study in Theoretical Science, National Taiwan University, Taipei 10607, Taiwan}

\date{\today}   

\begin{abstract}
We perform large-scale Monte Carlo simulations of the classical XY model on a three-dimensional  $L\times L \times L$  cubic lattice using the graphics processing unit (GPU). By the combination of Metropolis single-spin flip,  over-relaxation and parallel-tempering methods, we simulate systems up to $L=160$. Performing the finite-size scaling analysis,  we obtain  estimates of the critical exponents for the three-dimensional XY universality class: $\alpha=-0.01293(48)$ and $\nu=0.67098(16)$. Our estimate for the correlation-length exponent $\nu$, in contrast to previous theoretical estimates, agrees with the most recent experimental estimate $\nu_{\rm exp}=0.6709(1)$ at the superfluid transition of $^4$He in a microgravity environment.  
\end{abstract}

\maketitle
\end{CJK}

\section{Introduction}
One of the most beautiful ideas in physics is the renormalization-group (RG) theory,\cite{Wilson1975} which states that  near a critical point, the nature of the phase transition can be described by a few universal properties. These universal properties depend only upon the spatial dimensionality,  and the symmetry of the order parameter, regardless of the microscopic details of the system.  This indicates that the phase transitions can be classified into different \textit{universality classes}, and  the asymptotic critical behaviors of each class are described by a set of critical exponents and scaling functions. Among these, 
the three-dimensional (3D)  XY or O(2) universality class is the most extensively studied due to its relevance to the nature of the phase transitions in several physical systems, such as the $\lambda$-transition in  $^4$He. Experimentally, this superfluid transition permits the most accurate measurements of the critical exponents up to date in the micro-gravity environment.\cite{Lipa2003,*Lipa1996} The most recent value of  the correlation-length exponent is $\nu_{\rm exp}=0.6709(1)$, which is derived  from the measured value of the specific-heat exponent $\alpha$ via the hyperscaling relation.\cite{Lipa2003} 

\begin{figure}[bt]
 \includegraphics[width=\columnwidth]{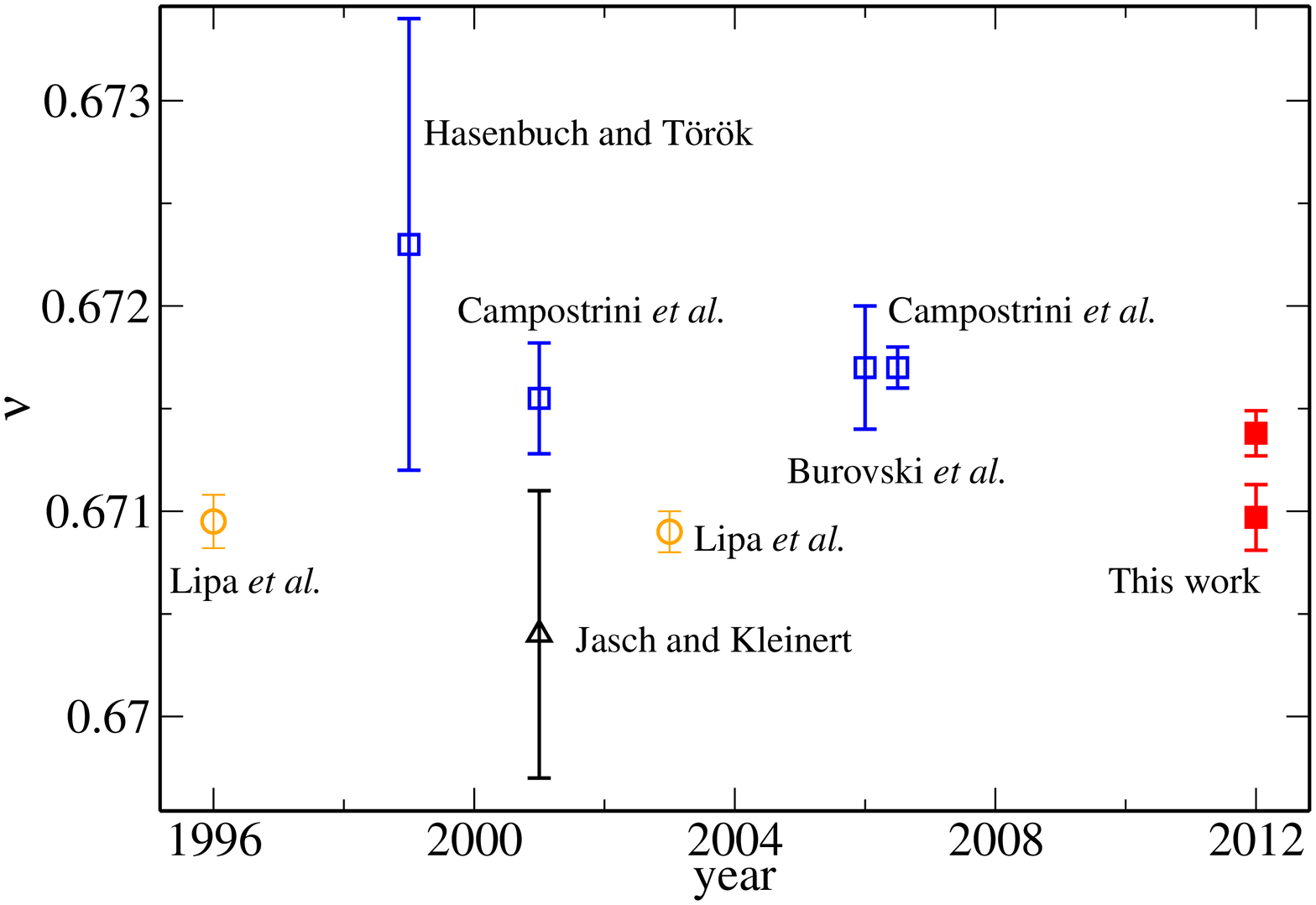} 
 \caption{ (Color online) Results for $\nu$ as a function of time. The circles show the experimental values, the upper triangle depicts the field-theoretical calculations, the squares show the Monte Carlo results, and the filled squares are the results of this work.}
 \label{nu_comparison}
\end{figure}

The XY universality class has been studied by various theoretical approaches: analytical field-theoretical methods\cite{Guillou1980, Jasch2001}, and numerical methods such as  high-temperature (HT) expansions,\cite{Campostrini2001} and Monte Carlo (MC) simulations.\cite{Li1989,Gottlob1993,Hasenbusch1999,Campostrini2001,Burovski2006, Campostrini2006}  
MC simulations combined with the finite-size scaling (FSS) technique\cite{Fisher1972}  have long been used to estimate  the critical exponents of phase transitions. With smaller system sizes, deviation from the universal behavior due to the irrelevant scaling operators can be the source of systematic errors in the FSS analyses, and corrections to scaling become necessary. It has been proposed that the corrections to scaling can be minimized by simulating the 3D two-component $\phi^4$ model on a simple cubic lattice, which  belongs to the 3D XY universality class, with a proper choice of a parameter in the model.\cite{Hasenbusch1999,Campostrini2001} However, the effect of sub-leading terms can only be partially suppressed due to the limited resolution in the tuning parameter. On the other hand, by including sub-leading corrections in the fits, it is argued that it is possible to  obtain more precise estimates of the critical exponents.\cite{Burovski2006} The history of recent results for the correlation-length exponent $\nu$ is given in Fig.~\ref{nu_comparison}.

Including the sub-leading correction brings complication since it is necessary to perform a high-dimensional fit to a non-polynomial function,  which might be sensitive to the numerical instability; therefore, a direct simulation of larger system sizes is desirable. Large-size simulations 
are crucial in developing a  clear signature of criticality, and for accurate determination of the critical properties.   In recent years, the advance of general purpose computing on the graphics processing units (GPUs) makes it possible to perform large-scale simulations in a massively parallel scheme.\cite{gpgpu}  In this paper, we present our MC simulations of the XY model on an $L\times L\times L$ cubic lattice  up to $L=160$ on GPU. Using  data obtained from  large-size systems, we are able to obtain the critical exponents with higher precision than previously achieved. 
This paper is organized as follows. In Sec.~\ref{sec:model}, we briefly discuss our simulation and analysis methods. Results of the simulation and a comparison with other works are presented in Sec.~\ref{sec:results}. Finally, we conclude in Sec.~\ref{sec:conclusion}.

\section{\label{sec:model}  Model and Method}
We simulate the classical XY model of the unit-length vectors $\mathbf{S}_i=(\cos \theta_i, \sin \theta_i)$ on an $L\times L\times L$ cubic lattice with the Hamiltonian,
\begin{equation}
\mathcal{H} = -J\sum_{\langle i, j \rangle} \mathbf{S}_i \cdot \mathbf{S}_j=-J\sum_{\langle i, j \rangle} \cos (\theta_i-\theta_j)
\end{equation}
where $\langle i,j \rangle$ indicates the nearest neighbor. The periodic boundary condition is applied.  In the following, we set $J = k_B =1$.  

\begin{figure}[tb]
\includegraphics[width=\columnwidth]{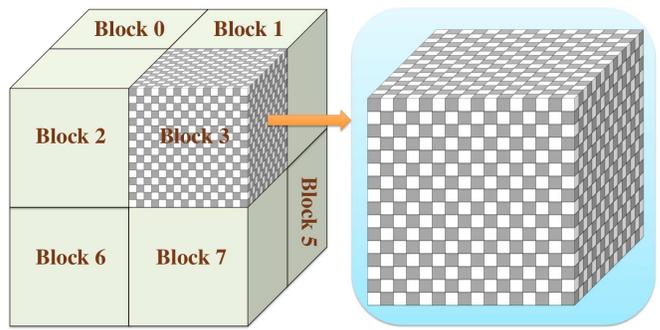}
 \caption{(Color online)  Checkerboard decomposition in 3D. The full lattice is decomposed into blocks containing $16\times 16\times 16=4096$ spins each. Each block is assigned to a thread block containing $ 8\times 8\times 8=512$ threads. Each thread manipulate  $2\times 2 \times 2=8$ spins, four black sites and four white sites.   } \label{fig:3d_blocks}
\end{figure}

We implement the GPU version of the Monte Carlo simulation 
based on the NVIDIA CUDA framework. \cite{CUDA} We  refer interested readers to available literature for
an introduction to the details of the GPU hardware and the programming models.\cite{ CUDA, *CUDAbook}
We implement three update schemes for the GPU version of the Monte Carlo simulation: parallel Metropolis single-spin flip,  over-relaxation, and parallel-tempering methods. One Monte Carlo step (MCS) is  defined as one Metropolis sweep and  one over-relaxation sweep  of the entire lattice, followed by one parallel-tempering exchange. To implement the parallel Metropolis and over-relaxation updates suitable for the GPU,
we divide the entire lattice into  blocks of  $16 \times 16 \times 16=4096$ spins.   Each block is  decomposed into sub-blocks which belong to two different sub-lattices (Fig.~\ref{fig:3d_blocks}). Each block is assigned to a \textit{thread block}\cite{CUDA,*CUDAbook} containing $ 8\times 8\times 8=512$ threads, which execute the same GPU \textit{kernel} in parallel.\cite{CUDA,*CUDAbook}   Each thread  is responsible for updating $2\times 2 \times 2=8$ spins, with four black sites and four white sites,  so that there are enough arithmetic operations   to hide the latency of the global memory accesses.\cite{CUDA, *CUDAbook} We  apply the  checkerboard decomposition algorithm to perform the Metropolis single-spin flips in parallel.\cite{Preis2009,Weigel2011} We first update all the black spins in parallel via a GPU kernel. After all the black spins belonging to different blocks are updated,  another kernel is launched to  update all  the white spins. Due to the special architecture of the GPU,  the commonly used Mersenne-Twister (MT) random number generator (RNG) can not be efficiently implemented  at the thread level. Instead,  we use a faster RNG implementation specially designed  for the GPU architecture, the \textit{Warp Generator}.~\cite{WarpGenerator} We note that although it has a smaller period of $2^{1024}-1$ than MT ($2^{19937}-1$), we do not find any noticeable statistical bias compared with the CPU runs using MT.

It is well established that the single-spin flip  Metropolis update suffers from the critical slowing down near the critical point and one has to resort to cluster updates.\cite{Swendsen1987, Wolff1989} However, implementation of the cluster update  on GPU is complicated and less efficient.\cite{Weigel2011b} We instead implement the microcanonical over-relaxation update.\cite{Creutz1987,Li1989}  The new value of the spin on site $i$ is obtained by  reflecting the spin at its local molecular field $\mathbf{H}_i=- \sum_{\langle ij\rangle} \mathbf{S}_j$, 
\begin{equation}
\mathbf{S}_i'=-\mathbf{S}_i+2\frac{\mathbf{S}_i \cdot \mathbf{H}_i}{H_i^2}\mathbf{H}_i
\end{equation}
This update  maps the system from a point in the phase space to another point with exactly the same energy. After several sweeps, the system is able to explore a larger region of the phase space without being stuck in a particular local minimum, and the simulation becomes ergodic.

To better equilibrate the simulation, we also perform the parallel-tempering (PT) Monte Carlo.\cite{Hukushima1996} In the PT scheme, many replicas at different temperatures are simulated simultaneously. After a certain number of MCSs, we swap two adjacent configurations $X_m, X_n$ at neighboring temperatures $T_m, T_n$  with the acceptance probability of
\begin{equation}
W(X_m, T_m | X_n, T_n) = {\rm min} \left[ 1,e^{(1/T_m - 1/T_n)(E_m - E_n)} \right],
\end{equation}
where  $E_n$ is the total energy of replica $n$.

We measure the following quantities during the simulation: magnetization $m$, Binder ratio $U_L$ and spin stiffness $\rho_s$ with the following estimators,
\begin{eqnarray}
m   &=& \frac{1}{N}\left\langle \sqrt{ \left(\sum_j \sin \theta_j\right)^2+\left(\sum_j \cos \theta_j\right)^2} \right\rangle,\\
U_L &=& 1 - \frac{\left\langle m^4\right\rangle}{3\left\langle m^2\right\rangle^2},\\
 \rho_s &=& \frac{1}{3N} \sum_{\hat{\mu}} \Bigg[  \left\langle \sum_{\langle i,j\rangle} \cos{(\theta_i-\theta_j)} (\hat{\epsilon}_{ij} \cdot \hat{\mu})^2  \right\rangle \nonumber \\
 && - \frac{1}{T}  \left\langle \left( \sum_{\langle i,j\rangle} \sin{(\theta_i-\theta_j)} (\hat{\epsilon}_{ij} \cdot 
 \hat{\mu}) \right)^2  \right\rangle \Bigg],
\end{eqnarray}
where $\hat{\mu} = \hat{x}, \hat{y}$ and $\hat{z}$, and $\hat{\epsilon}_{ij}$ is the unit vector connecting nearest-neighbor sites $i$ and $j$. 

To reduce the amount of data transfer between the CPU and the GPU, we store all the spin configurations at  different temperatures in the GPU global memory, and all updates are performed through the kernel functions on the GPU. Measurements are also performed on the GPU and the results are sent back to the CPU for binning. 
Simulations are carried out at 34 temperatures ranging from $T = 2.1$ to $T = 2.3$ for $L = 64, 80, 96,$ and $128$ and at 13 temperatures ranging from $T = 2.19$ to $T = 2.21$  for  $L = 160$. The temperature set is chosen such that the acceptance rate of swaps is independent of the temperatures. After $3\times10^6$ MCSs for equilibrium, $1.3\times10^7$ measurements are made for $L = 64, 80,$ and $96$ and $2\times10^7$ measurements are made for $L = 128$ and $160$. The data are blocked into several bins, each consisting of $10^5$ measurements, for further analyses.  Error bars are given by one standard deviation.
The simulations were performed on Nvidia Tesla M2090, and  took approximately 110 days of GPU time in total to accumulate the whole data set.

\begin{table*}[t] \caption{Comparison of the critical exponents determined via various methods for three-dimensional XY universality class. The quantities with asterisk are calculated using the scaling relation $\gamma = (2-\eta)\nu$ or the hyperscaling relation [Eq.~(\ref{hyper})], and errors are calculated by treating variables as independent.}
\label{comparison}
\begin{ruledtabular}
\begin{tabular}{lllllll}
\textrm{Method} & \textrm{Ref.} & $\alpha$ & $\beta$ & $\gamma$ & $\nu$\\
\colrule
MC & this work & -0.01293(48)$^{*}$ & - & - & 0.67098(16) \\
 &  & -0.01414(33)$^{*}$ & 0.34910(12) & 1.31594(41)$^{*}$ & 0.67138(11) \\
MC + IHT\footnote{improved high-temperature expansion} & [\onlinecite{Campostrini2006}] (2006) & -0.0151(3)$^{*}$&  0.3486(1)$^{*}$& 1.3178(2) & 0.6717(1)\\
MC  & [\onlinecite{Burovski2006}] (2006) & -0.0151(9)$^{*}$ & - & - & 0.6717(3)\\
MC+IHT & [\onlinecite{Campostrini2001}] (2001) & -0.0146(8)$^{*}$ & 0.3485(3)$^{*}$ & 1.3177(5) & 0.67155(27)\\
MC & [\onlinecite{Campostrini2001}] (2001) & -0.0148(15)$^{*}$ & 0.3485(2)$^{*}$ & 1.3177(10)$^{*}$ & 0.6716(5)\\
$\phi^4$ \footnote{two-component $\phi^4$ field theory} & [\onlinecite{Jasch2001}] (2001) & -0.0112(21) & 0.3474(11)$^{*}$ & 1.3164(8) & 0.6704(7)\\
MC & [\onlinecite{Hasenbusch1999}] (1999) & -0.0169(33)$^{*}$ & 0.349(2)$^{*}$ & 1.3190(24) & 0.6723(11)\\
PRG\footnote{phenomenological RG} & [\onlinecite{Gottlob1993}] (1993) & -0.014(21)$^{*}$ & - & 1.307(14)$^{*}$ & 0.662(7)\\
MC & [\onlinecite{Gottlob1993}] (1993) & - & - & 1.324(1) & - \\
exp\footnote{experiment of $^{4}$He superfluid} & [\onlinecite{Lipa2003}] (2003) & -0.0127(3) & - & - & 0.6709(1)\\
exp & [\onlinecite{Lipa1996}] (1996) & -0.01056(38) & - & - & 0.67019(13)\\
\end{tabular}
\end{ruledtabular}
\end{table*}

\section{\label{sec:results}Results}
We perform the finite-size scaling analyses to extract the critical behaviors in the thermodynamic limit.\cite{Fisher1972,Binder1981}
The singular part of the free energy with critical exponent $\kappa$ in zero-field can be describe by the scaling ansatz
\begin{equation}
F(t,L) = L^{\kappa/\nu}\mathcal{F}^0(tL^{1/\nu}), 
\end{equation}
where  $\nu$ is the correlation-length critical exponent, $t = (T-T_c)/T_c$ is  the reduced temperature , and $\mathcal{F}^0(x)$ is an universal function which is analytic as $x\rightarrow 0$.

\begin{figure}[tb]
 \includegraphics[width=\columnwidth,clip]{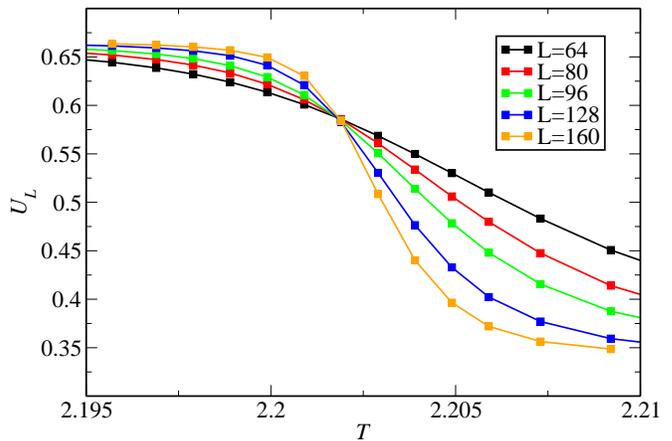} 
 \caption{(Color online) $U_L$ vs temperature for $L = $ 64, 80, 96, 128 and 160. The error bars are smaller than the symbols.}
 \label{binder}
\end{figure}

We first use the spin stiffness $\rho_s$ and the Binder cumulant $U_L$ to estimate $T_c$. 
The spin stiffness $\rho_s$ scales as 
\begin{equation}
\rho_s= L^{-1}\mathcal{H}^0(tL^{1/\nu})
\end{equation} for the 3D XY model. This indicates that $\rho_s L$ at different sizes should intersect at $T_c$ (Fig.~\ref{rhoL}). 
Also, the Binder cumulant $U_L$   for different sizes  also intersect at $T_c$ (Fig.~\ref{binder}). In both cases, we obtain the crossing of the curves at a consistent $T_c= 2.2019$.

\begin{figure}[tb]
 \includegraphics[width=\columnwidth,clip]{rhoL_vs_T.eps} 
 \caption{(Color online) $\rho_sL$ vs temperature for $L = $ 64, 80, 96, 128 and 160. The error bars are smaller than the symbols.}
 \label{rhoL}
\end{figure}

\begin{figure}[tb]
 \includegraphics[width=\columnwidth, clip]{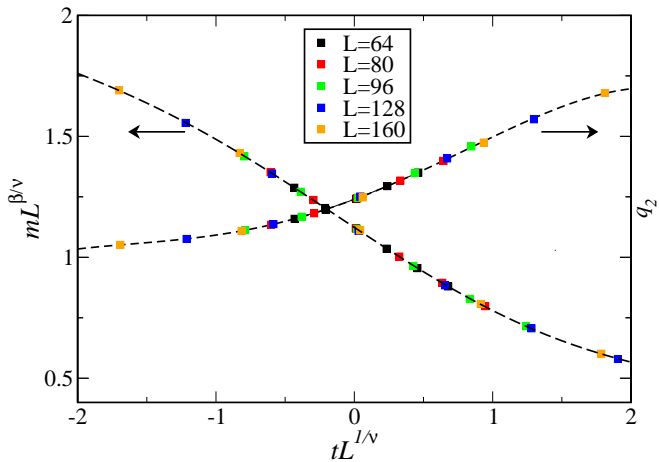}
 \caption{(Color online) Scaling of $m$ and $q_2$ for $L$=64 (black), 80 (red), 96 (green), 128 (blue) and 160 (orange). The dashed lines correspond to polynomial fits to the corresponding scaling functions.}
 \label{collapse}
\end{figure}

Further refinement of the analysis is carried out by data collapse. 
We first perform data collapse on the magnetization $m$ (Fig. \ref{collapse}) to obtain $T_c$, $\beta$ and $\nu$. 
Near $T_c$, $m$ has the scaling form 
\begin{equation}
m(t,L) = L^{-\beta/\nu}\mathcal{M}^0(tL^{1/\nu}),\label{mag}
\end{equation} 
where $\mathcal{M}^0$ is an universal function. 
We use bootstrap resampling technique\cite{Wu1986}  to decorrelate the data obtained by the PT at different temperatures.  
In each resampling, 1000 values are randomly chosen from the 130 (or 200) measurements and then take the average. The resampling is repeated for 130 (or 200) times to generate a new data set. This data set is used to perform data collapse to obtain $T_c$, $\beta$ and  $\nu$. Temperatures between 2.1990 and 2.2050 are used for data collapse. A fifth-order polynomial is used to fit the scaling function $\mathcal{M}^0$, and the phase space of $T_c$, $\beta$ and $\nu$ is scanned for the best collapse, where the reduced chi-square $\chi_{\rm red}=\chi/{\rm d.o.f.}$ approaches one. The procedure is  repeated  100 times to estimate the error bars of $T_c$, $\beta$ and $\nu$ (Fig. \ref{collapse}). Our estimates are $T_c = 2.201852(1)$, $\beta = 0.34910(12)$ and $\nu = 0.67138(11)$ with the average  $\chi_{\rm red}^2 \approx 1.2938$.  We  use the hyperscaling relation\cite{Fisher1998}
\begin{equation} \label{hyper}
2 - \alpha = 3 \nu = 2 \beta + \gamma, 
\end{equation}
to obtain the estimates for $\alpha=-0.01414(33)$ and $\gamma=1.31594(41)$.

We also perform data collapse on $q_2$ defined as,  
\begin{equation} \label{q2}
Q_2(t,L) = 3(1-U_L) = \frac{\langle m^4\rangle}{\langle m^2\rangle^2} = q_2(tL^{1/\nu}).
\end{equation}
The same procedure of resampling is applied as above. 
Data at temperatures between 2.1990 and 2.2040 are used for data collapse. A fourth-order polynomial is used to fit $q_2$. The phase space of $T_c$ and $\nu$ is scanned to produce the best collapse. The procedure is  repeated  100 times to estimate the error bars for $T_c$ and $\nu$. Figure~\ref{collapse} also shows the scaling plot of $q_2$ for $L=64,80,96, 128$ and 160. Our estimates give $T_c = 2.2018312(6)$, $\nu = 0.67098(16)$ with average $\chi_{red}^2 \approx 1.2458$. Using the hyperscaling relation Eq.~(\ref{hyper}), we obtain the estimates for $\alpha=-0.01293(48)$.

In  Table~\ref{comparison}, we compare our estimates with recent results for the critical exponents of the 3D XY universality class. Our  estimates of $\nu$ are consistent with each other within two standard deviations (Fig.~\ref{nu_comparison}), and  the error bars  are comparable with the experimental precision.  However, they are smaller than previous theoretical estimates for $\nu$,\cite{Burovski2006,Campostrini2006} and, contrary to previous claims, are consistent with the experimental estimate\cite{Lipa2003}. This might be attributed to the largest system sizes ($L=128, 160$)  with high statistics that we can simulate, and a better finite-size scaling analysis can be performed.  

\section{\label{sec:conclusion}Conclusion}
We  perform large-scale Monte Carlo calculations of the 3D XY model. Implementation of efficient, highly parallel  Monte Carlo update schemes on GPU  enables us to perform simulations on lattices up to $L$ = 160. With larger system sizes, we are able to perform finite-size scaling and obtain a five-digit accuracy of the critical exponents in a significantly less amount of computation time. With the current accuracy for $\nu$ and $\alpha$ in our simulations, contrary to previous theoretical studies, our results  suggest that the theoretical estimates of the critical exponents for the 3D XY universality class are consistent with experimental results within two standard deviations. This suggest that the $\lambda$-transition in He$^{4}$  can indeed be accurately described by the 3D XY universality class. It would be interesting to revisit the models studied previously to further confirm results obtained in this paper. 
\begin{acknowledgments}
We thank A. W. Sandvik for useful discussions. This work is partially supported by NSC in Taiwan through Grant No.
100-2112-M-002-013-MY3 (Y. D. H., Y.J. K.), and by NTU Grant numbers 101R891004 (Y.J.
K.). Travel support from NCTS in Taiwan is also acknowledged. 
\end{acknowledgments}

\bibliography{3dxy_model}

\providecommand{\noopsort}[1]{}\providecommand{\singleletter}[1]{#1}%
\begin{thebibliography}{26}%
\makeatletter
\providecommand \@ifxundefined [1]{%
 \@ifx{#1\undefined}
}%
\providecommand \@ifnum [1]{%
 \ifnum #1\expandafter \@firstoftwo
 \else \expandafter \@secondoftwo
 \fi
}%
\providecommand \@ifx [1]{%
 \ifx #1\expandafter \@firstoftwo
 \else \expandafter \@secondoftwo
 \fi
}%
\providecommand \natexlab [1]{#1}%
\providecommand \enquote  [1]{``#1''}%
\providecommand \bibnamefont  [1]{#1}%
\providecommand \bibfnamefont [1]{#1}%
\providecommand \citenamefont [1]{#1}%
\providecommand \href@noop [0]{\@secondoftwo}%
\providecommand \href [0]{\begingroup \@sanitize@url \@href}%
\providecommand \@href[1]{\@@startlink{#1}\@@href}%
\providecommand \@@href[1]{\endgroup#1\@@endlink}%
\providecommand \@sanitize@url [0]{\catcode `\\12\catcode `\$12\catcode
  `\&12\catcode `\#12\catcode `\^12\catcode `\_12\catcode `\%12\relax}%
\providecommand \@@startlink[1]{}%
\providecommand \@@endlink[0]{}%
\providecommand \url  [0]{\begingroup\@sanitize@url \@url }%
\providecommand \@url [1]{\endgroup\@href {#1}{\urlprefix }}%
\providecommand \urlprefix  [0]{URL }%
\providecommand \Eprint [0]{\href }%
\providecommand \doibase [0]{http://dx.doi.org/}%
\providecommand \selectlanguage [0]{\@gobble}%
\providecommand \bibinfo  [0]{\@secondoftwo}%
\providecommand \bibfield  [0]{\@secondoftwo}%
\providecommand \translation [1]{[#1]}%
\providecommand \BibitemOpen [0]{}%
\providecommand \bibitemStop [0]{}%
\providecommand \bibitemNoStop [0]{.\EOS\space}%
\providecommand \EOS [0]{\spacefactor3000\relax}%
\providecommand \BibitemShut  [1]{\csname bibitem#1\endcsname}%
\let\auto@bib@innerbib\@empty
\bibitem [{\citenamefont {Wilson}(1975)}]{Wilson1975}%
  \BibitemOpen
  \bibfield  {author} {\bibinfo {author} {\bibfnamefont {K.~G.}\ \bibnamefont
  {Wilson}},\ }\href {\doibase 10.1103/RevModPhys.47.773} {\bibfield  {journal}
  {\bibinfo  {journal} {Rev. Mod. Phys.}\ }\textbf {\bibinfo {volume} {47}},\
  \bibinfo {pages} {773} (\bibinfo {year} {1975})}\BibitemShut {NoStop}%
\bibitem [{\citenamefont {Lipa}\ \emph {et~al.}(2003)\citenamefont {Lipa},
  \citenamefont {Nissen}, \citenamefont {Stricker}, \citenamefont {Swanson},\
  and\ \citenamefont {Chui}}]{Lipa2003}%
  \BibitemOpen
  \bibfield  {author} {\bibinfo {author} {\bibfnamefont {J.~A.}\ \bibnamefont
  {Lipa}}, \bibinfo {author} {\bibfnamefont {J.~A.}\ \bibnamefont {Nissen}},
  \bibinfo {author} {\bibfnamefont {D.~A.}\ \bibnamefont {Stricker}}, \bibinfo
  {author} {\bibfnamefont {D.~R.}\ \bibnamefont {Swanson}}, \ and\ \bibinfo
  {author} {\bibfnamefont {T.~C.~P.}\ \bibnamefont {Chui}},\ }\href@noop {}
  {\bibfield  {journal} {\bibinfo  {journal} {Phys. Rev. B}\ }\textbf {\bibinfo
  {volume} {68}},\ \bibinfo {pages} {174518} (\bibinfo {year}
  {2003})}\BibitemShut {NoStop}%
\bibitem [{\citenamefont {Lipa}\ \emph {et~al.}(1996)\citenamefont {Lipa},
  \citenamefont {Swanson}, \citenamefont {Nissen}, \citenamefont {Chui},\ and\
  \citenamefont {Israelsson}}]{Lipa1996}%
  \BibitemOpen
  \bibfield  {author} {\bibinfo {author} {\bibfnamefont {J.~A.}\ \bibnamefont
  {Lipa}}, \bibinfo {author} {\bibfnamefont {D.~R.}\ \bibnamefont {Swanson}},
  \bibinfo {author} {\bibfnamefont {J.~A.}\ \bibnamefont {Nissen}}, \bibinfo
  {author} {\bibfnamefont {T.~C.~P.}\ \bibnamefont {Chui}}, \ and\ \bibinfo
  {author} {\bibfnamefont {U.~E.}\ \bibnamefont {Israelsson}},\ }\href@noop {}
  {\bibfield  {journal} {\bibinfo  {journal} {Phys. Rev. Lett.}\ }\textbf
  {\bibinfo {volume} {76}},\ \bibinfo {pages} {944} (\bibinfo {year}
  {1996})}\BibitemShut {NoStop}%
\bibitem [{\citenamefont {Guillou}\ and\ \citenamefont
  {Zinn-Justin}(1980)}]{Guillou1980}%
  \BibitemOpen
  \bibfield  {author} {\bibinfo {author} {\bibfnamefont {J.~C.~L.}\
  \bibnamefont {Guillou}}\ and\ \bibinfo {author} {\bibfnamefont
  {J.}~\bibnamefont {Zinn-Justin}},\ }\href@noop {} {\bibfield  {journal}
  {\bibinfo  {journal} {Phys. Rev. B}\ }\textbf {\bibinfo {volume} {21}},\
  \bibinfo {pages} {3976} (\bibinfo {year} {1980})}\BibitemShut {NoStop}%
\bibitem [{\citenamefont {Florian~Jasch}(2001)}]{Jasch2001}%
  \BibitemOpen
  \bibfield  {author} {\bibinfo {author} {\bibfnamefont {H.~K.}\ \bibnamefont
  {Florian~Jasch}},\ }\href@noop {} {\bibfield  {journal} {\bibinfo  {journal}
  {J. Math. Phys.}\ }\textbf {\bibinfo {volume} {42}},\ \bibinfo {pages} {52}
  (\bibinfo {year} {2001})}\BibitemShut {NoStop}%
\bibitem [{\citenamefont {Campostrini}\ \emph {et~al.}(2001)\citenamefont
  {Campostrini}, \citenamefont {Hasenbusch}, \citenamefont {Pelissetto},
  \citenamefont {Rossi},\ and\ \citenamefont {Vicari}}]{Campostrini2001}%
  \BibitemOpen
  \bibfield  {author} {\bibinfo {author} {\bibfnamefont {M.}~\bibnamefont
  {Campostrini}}, \bibinfo {author} {\bibfnamefont {M.}~\bibnamefont
  {Hasenbusch}}, \bibinfo {author} {\bibfnamefont {A.}~\bibnamefont
  {Pelissetto}}, \bibinfo {author} {\bibfnamefont {P.}~\bibnamefont {Rossi}}, \
  and\ \bibinfo {author} {\bibfnamefont {E.}~\bibnamefont {Vicari}},\
  }\href@noop {} {\bibfield  {journal} {\bibinfo  {journal} {Phys. Rev. B}\
  }\textbf {\bibinfo {volume} {63}},\ \bibinfo {pages} {214503} (\bibinfo
  {year} {2001})}\BibitemShut {NoStop}%
\bibitem [{\citenamefont {Li}\ and\ \citenamefont {Teitel}(1989)}]{Li1989}%
  \BibitemOpen
  \bibfield  {author} {\bibinfo {author} {\bibfnamefont {Y.~H.}\ \bibnamefont
  {Li}}\ and\ \bibinfo {author} {\bibfnamefont {S.}~\bibnamefont {Teitel}},\
  }\href@noop {} {\bibfield  {journal} {\bibinfo  {journal} {Phys. Rev. B}\
  }\textbf {\bibinfo {volume} {40}},\ \bibinfo {pages} {9122} (\bibinfo {year}
  {1989})}\BibitemShut {NoStop}%
\bibitem [{\citenamefont {Gottlob}\ and\ \citenamefont
  {Hasenpusch}(1993)}]{Gottlob1993}%
  \BibitemOpen
  \bibfield  {author} {\bibinfo {author} {\bibfnamefont {A.~P.}\ \bibnamefont
  {Gottlob}}\ and\ \bibinfo {author} {\bibfnamefont {M.}~\bibnamefont
  {Hasenpusch}},\ }\href@noop {} {\bibfield  {journal} {\bibinfo  {journal}
  {Physica A}\ }\textbf {\bibinfo {volume} {201}},\ \bibinfo {pages} {593}
  (\bibinfo {year} {1993})}\BibitemShut {NoStop}%
\bibitem [{\citenamefont {Hasenbusch}\ and\ \citenamefont
  {T\"{o}r\"{o}k}(1999)}]{Hasenbusch1999}%
  \BibitemOpen
  \bibfield  {author} {\bibinfo {author} {\bibfnamefont {M.}~\bibnamefont
  {Hasenbusch}}\ and\ \bibinfo {author} {\bibfnamefont {T.}~\bibnamefont
  {T\"{o}r\"{o}k}},\ }\href@noop {} {\bibfield  {journal} {\bibinfo  {journal}
  {J. Phys. A: Math. Gen}\ }\textbf {\bibinfo {volume} {32}},\ \bibinfo {pages}
  {6361} (\bibinfo {year} {1999})}\BibitemShut {NoStop}%
\bibitem [{\citenamefont {Burovski}\ \emph {et~al.}(2006)\citenamefont
  {Burovski}, \citenamefont {Machta}, \citenamefont {Prokof'ev},\ and\
  \citenamefont {Svistunov}}]{Burovski2006}%
  \BibitemOpen
  \bibfield  {author} {\bibinfo {author} {\bibfnamefont {E.}~\bibnamefont
  {Burovski}}, \bibinfo {author} {\bibfnamefont {J.}~\bibnamefont {Machta}},
  \bibinfo {author} {\bibfnamefont {N.}~\bibnamefont {Prokof'ev}}, \ and\
  \bibinfo {author} {\bibfnamefont {B.}~\bibnamefont {Svistunov}},\ }\href@noop
  {} {\bibfield  {journal} {\bibinfo  {journal} {Phys. Rev. B.}\ }\textbf
  {\bibinfo {volume} {74}},\ \bibinfo {pages} {132502} (\bibinfo {year}
  {2006})}\BibitemShut {NoStop}%
\bibitem [{\citenamefont {Campostrini}\ \emph {et~al.}(2006)\citenamefont
  {Campostrini}, \citenamefont {Hasenbusch}, \citenamefont {Pelissetto},\ and\
  \citenamefont {Vicari}}]{Campostrini2006}%
  \BibitemOpen
  \bibfield  {author} {\bibinfo {author} {\bibfnamefont {M.}~\bibnamefont
  {Campostrini}}, \bibinfo {author} {\bibfnamefont {M.}~\bibnamefont
  {Hasenbusch}}, \bibinfo {author} {\bibfnamefont {A.}~\bibnamefont
  {Pelissetto}}, \ and\ \bibinfo {author} {\bibfnamefont {E.}~\bibnamefont
  {Vicari}},\ }\href {\doibase 10.1103/PhysRevB.74.144506} {\bibfield
  {journal} {\bibinfo  {journal} {Phys. Rev. B}\ }\textbf {\bibinfo {volume}
  {74}},\ \bibinfo {pages} {144506} (\bibinfo {year} {2006})}\BibitemShut
  {NoStop}%
\bibitem [{\citenamefont {Fisher}\ and\ \citenamefont
  {Barber}(1972)}]{Fisher1972}%
  \BibitemOpen
  \bibfield  {author} {\bibinfo {author} {\bibfnamefont {M.~E.}\ \bibnamefont
  {Fisher}}\ and\ \bibinfo {author} {\bibfnamefont {M.~N.}\ \bibnamefont
  {Barber}},\ }\href@noop {} {\bibfield  {journal} {\bibinfo  {journal} {Phys.
  Rev. Lett.}\ }\textbf {\bibinfo {volume} {28}},\ \bibinfo {pages} {1516}
  (\bibinfo {year} {1972})}\BibitemShut {NoStop}%
\bibitem [{gpg()}]{gpgpu}%
  \BibitemOpen
  \href@noop {} {}\bibinfo {note} {See \url{http://gpgpu.org} for articles and
  discussion of general-purpose computing with GPUs.}\BibitemShut {Stop}%
\bibitem [{CUD()}]{CUDA}%
  \BibitemOpen
  \href@noop {} {}\bibinfo {note}
  {\url{http://www.nvidia.com/object/cuda_home_new.html}}\BibitemShut {NoStop}%
\bibitem [{\citenamefont {Kirk}\ and\ \citenamefont {Hwu}(2010)}]{CUDAbook}%
  \BibitemOpen
  \bibfield  {author} {\bibinfo {author} {\bibfnamefont {D.~B.}\ \bibnamefont
  {Kirk}}\ and\ \bibinfo {author} {\bibfnamefont {W.~W.}\ \bibnamefont {Hwu}},\
  }\href@noop {} {\emph {\bibinfo {title} {Programming Massively Parallel
  Processors}}}\ (\bibinfo  {publisher} {Elsevier},\ \bibinfo {address}
  {Amsterdam},\ \bibinfo {year} {2010})\BibitemShut {NoStop}%
\bibitem [{\citenamefont {Preis}\ \emph {et~al.}(2009)\citenamefont {Preis},
  \citenamefont {Virnau}, \citenamefont {Paul},\ and\ \citenamefont
  {Schneider}}]{Preis2009}%
  \BibitemOpen
  \bibfield  {author} {\bibinfo {author} {\bibfnamefont {T.}~\bibnamefont
  {Preis}}, \bibinfo {author} {\bibfnamefont {P.}~\bibnamefont {Virnau}},
  \bibinfo {author} {\bibfnamefont {W.}~\bibnamefont {Paul}}, \ and\ \bibinfo
  {author} {\bibfnamefont {J.~J.}\ \bibnamefont {Schneider}},\ }\href {\doibase
  10.1016/j.jcp.2009.03.018} {\bibfield  {journal} {\bibinfo  {journal}
  {Journal of Computational Physics}\ }\textbf {\bibinfo {volume} {228}},\
  \bibinfo {pages} {4468 } (\bibinfo {year} {2009})}\BibitemShut {NoStop}%
\bibitem [{\citenamefont {Weigel}(2011{\natexlab{a}})}]{Weigel2011}%
  \BibitemOpen
  \bibfield  {author} {\bibinfo {author} {\bibfnamefont {M.}~\bibnamefont
  {Weigel}},\ }\href@noop {} {\bibfield  {journal} {\bibinfo  {journal}
  {Computer Physics Communications}\ }\textbf {\bibinfo {volume} {182}},\
  \bibinfo {pages} {1833} (\bibinfo {year} {2011}{\natexlab{a}})}\BibitemShut
  {NoStop}%
\bibitem [{War()}]{WarpGenerator}%
  \BibitemOpen
  \href@noop {} {}\bibinfo {note}
  {\url{http://cas.ee.ic.ac.uk/people/dt10/research/rngs-gpu-warp_generator.html}}\BibitemShut
  {NoStop}%
\bibitem [{\citenamefont {Swendsen}\ and\ \citenamefont
  {Wang}(1987)}]{Swendsen1987}%
  \BibitemOpen
  \bibfield  {author} {\bibinfo {author} {\bibfnamefont {R.~H.}\ \bibnamefont
  {Swendsen}}\ and\ \bibinfo {author} {\bibfnamefont {J.-S.}\ \bibnamefont
  {Wang}},\ }\href {\doibase 10.1103/PhysRevLett.58.86} {\bibfield  {journal}
  {\bibinfo  {journal} {Phys. Rev. Lett.}\ }\textbf {\bibinfo {volume} {58}},\
  \bibinfo {pages} {86} (\bibinfo {year} {1987})}\BibitemShut {NoStop}%
\bibitem [{\citenamefont {Wolff}(1989)}]{Wolff1989}%
  \BibitemOpen
  \bibfield  {author} {\bibinfo {author} {\bibfnamefont {U.}~\bibnamefont
  {Wolff}},\ }\href {\doibase 10.1103/PhysRevLett.62.361} {\bibfield  {journal}
  {\bibinfo  {journal} {Phys. Rev. Lett.}\ }\textbf {\bibinfo {volume} {62}},\
  \bibinfo {pages} {361} (\bibinfo {year} {1989})}\BibitemShut {NoStop}%
\bibitem [{\citenamefont {Weigel}(2011{\natexlab{b}})}]{Weigel2011b}%
  \BibitemOpen
  \bibfield  {author} {\bibinfo {author} {\bibfnamefont {M.}~\bibnamefont
  {Weigel}},\ }\href@noop {} {\enquote {\bibinfo {title} {Connected component
  identification and cluster update on gpu},}\ } (\bibinfo {year}
  {2011}{\natexlab{b}}),\ \Eprint {http://arxiv.org/abs/1105.5804v2}
  {arXiv:1105.5804v2} \BibitemShut {NoStop}%
\bibitem [{\citenamefont {Creutz}(1987)}]{Creutz1987}%
  \BibitemOpen
  \bibfield  {author} {\bibinfo {author} {\bibfnamefont {M.}~\bibnamefont
  {Creutz}},\ }\href@noop {} {\bibfield  {journal} {\bibinfo  {journal} {Phys.
  Rev. D}\ }\textbf {\bibinfo {volume} {36}},\ \bibinfo {pages} {515} (\bibinfo
  {year} {1987})}\BibitemShut {NoStop}%
\bibitem [{\citenamefont {Hukushima}\ and\ \citenamefont
  {Nemoto}(1996)}]{Hukushima1996}%
  \BibitemOpen
  \bibfield  {author} {\bibinfo {author} {\bibfnamefont {K.}~\bibnamefont
  {Hukushima}}\ and\ \bibinfo {author} {\bibfnamefont {K.}~\bibnamefont
  {Nemoto}},\ }\href@noop {} {\bibfield  {journal} {\bibinfo  {journal} {J.
  Phys. Soc. Jpn.}\ }\textbf {\bibinfo {volume} {65}},\ \bibinfo {pages} {1604}
  (\bibinfo {year} {1996})}\BibitemShut {NoStop}%
\bibitem [{\citenamefont {Binder}(1981)}]{Binder1981}%
  \BibitemOpen
  \bibfield  {author} {\bibinfo {author} {\bibfnamefont {K.}~\bibnamefont
  {Binder}},\ }\href@noop {} {\bibfield  {journal} {\bibinfo  {journal} {Phys.
  Rev. Lett.}\ }\textbf {\bibinfo {volume} {47}},\ \bibinfo {pages} {693}
  (\bibinfo {year} {1981})}\BibitemShut {NoStop}%
\bibitem [{\citenamefont {Wu}(1986)}]{Wu1986}%
  \BibitemOpen
  \bibfield  {author} {\bibinfo {author} {\bibfnamefont {C.~F.~J.}\
  \bibnamefont {Wu}},\ }\href@noop {} {\bibfield  {journal} {\bibinfo
  {journal} {Annals of Statistics}\ }\textbf {\bibinfo {volume} {14}},\
  \bibinfo {pages} {1261} (\bibinfo {year} {1986})}\BibitemShut {NoStop}%
\bibitem [{\citenamefont {Fisher}(1998)}]{Fisher1998}%
  \BibitemOpen
  \bibfield  {author} {\bibinfo {author} {\bibfnamefont {M.~E.}\ \bibnamefont
  {Fisher}},\ }\href@noop {} {\bibfield  {journal} {\bibinfo  {journal} {Rev.
  Mod. Phys.}\ }\textbf {\bibinfo {volume} {70}},\ \bibinfo {pages} {653}
  (\bibinfo {year} {1998})}\BibitemShut {NoStop}%
\end{thebibliography}%

\end{document}